# Direct Evidence for the Localized Single-Triplet Excitations and the Dispersive Multi-Triplets Excitations in SrCu$_2$(BO$_3$)$_2$

H. Kageyama[1, *], M. Nishi[2], N. Aso[2], K. Onizuka[1], T. Yosihama[2], K. Nukui[2], K. Kodama[3], K. Kakurai[2], and Y. Ueda[1]

[1] *Material Design and Characterization Laboratory, Institute for Solid State Physics, University of Tokyo, Kashiwanoha, Kashiwa, Chiba 277-8581, Japan*

[2] *Neutron Scattering Laboratory, Institute for Solid State Physics, University of Tokyo, 106-1 Shirakata, Tokai, Ibaraki 319-1106, Japan*

[3] *Division of New Materials Science, Institute for Solid State Physics, University of Tokyo, Roppongi, Kashiwanoha, Kashiwa, Chiba 277-8581, Japan*

ABSTRACT

We performed inelastic neutron scattering on the 2D Shastry-Sutherland system SrCu$_2$($^{11}$BO$_3$)$_2$ with an exact dimer ground state. Three energy levels at around 3, 5 and 9 meV were observed at 1.7 K. The lowest excitation at 3.0 meV is almost dispersionless with a bandwidth of 0.2 meV at most, showing a significant constraint on a single-triplet hopping owing to the orthogonality of the neighboring dimers. In contrast, the correlated two-triplets excitations at 5 meV exhibit a more dispersive behavior.

Recently, much experimental and theoretical effort has been spent to understand the quantum spin dynamics in low-dimensional magnetic systems. The interest has been mainly revived by the discovery of two-dimensional (2D) $S=1/2$ spin correlations in high-$T_c$ superconducting cuprates. In course of these studies, many new low-dimensional magnetic systems with a spin-singlet ground state have been discovered, e.g., Haldane, spin-Peierls, ladder, plaquette systems, to name only few [1]. The spin dynamics of these systems show a spin excitation gap from the non-magnetic singlet ground state to the first excited triplet state, which may be regarded as a direct consequence of the quantum spin fluctuations. However, little is known about the frustration effect in these extreme quantum spin system. It is, e.g., still controversial how big a role the finite next-nearest interaction in the spin-Peierls compound $CuGeO_3$ plays in establishing the observed singlet ground state.

$SrCu_2(BO_3)_2$ has a tetragonal unit cell with lattice parameters $a=b=8.995$ Å and $c=6.649$ Å at room temperature. It consists of alternately stacked Sr- and $CuBO_3$-planes. A neighboring pair of planar rectangular $CuO_4$ forms a spin dimer and the dimers are connected orthogonally by a triangular $BO_3$. It was pointed out by Miyahara and Ueda [2] that $SrCu_2(BO_3)_2$ can be mapped on the Shastry and Sutherland model [3], which has an exactly solvable ground state. They have also shown that a phase transition from a Néel to singlet dimer state occurs at the critical ratio of interdimer ($J'$) and intradimer ($J$) exchange of $(J'/J)_c=0.70$ (or 0.691 [4]) with decreasing $J'/J$. Based on this theoretical model, it has been established from our previous experiments [5-10] that $SrCu_2(BO_3)_2$ is an $S=1/2$ highly frustrated spin gap system with an energy gap of  =34 K and with a ratio of $J'/J=0.68$ ($J=100$ K, $J'=68$ K) very close to the critical value. The exactness of the ground state allows theorists to investigate its excited states in details. One of the most remarkable results is that a *single* triplet can propagate only from the sixth order in the perturbation calculations, indicating the *significant constraint on the hopping* as shown by Miyahara and Ueda [2]. In addition, the fifteenth-order calculation of the dispersion relation of the *single* triplet by Weihong *et al.* revealed a prominent $J'/J$-dependence of the bandwidth [4].

So far no direct experimental investigation has been carried out to address the questions as whether the *single* triplet excitation is actually localized and if so how difficult its hopping process is. The appearance of the fractional magnetization plateaux of the total magnetization [5] is associated with the localization of the *multi*-triplets and also with the *commensurability* to the lattice, which is a kind of the Mott transition in the broad sense of the word. But, the magnetization measurement only provides the *static* information of the excited triplets, and the applied magnetic field splits the triplet state into three levels; $S_z$=-1, 0, +1 (Zeeman splitting), making the situation different from that considered by Miyahara and Ueda. Likewise, the dynamical properties of the two-triplets mode has not been investigated yet—although this mode has been detected by ESR [7] and Raman scattering [8], both methods can probe the excitation only at $Q=(0, 0)$.

It is thus interesting and important to probe directly the wave vector dependence of spin excitations in $SrCu_2(BO_3)_2$, the Shastry-Sutherland model with $J'/J=0.68$ close to the critical boundary, as the prototype of strongly frustrated quantum spin system. In this Letter, we report the first inelastic neutron scattering experiments in bulk single crystal to investigate the *dynamical* properties of the triplet excitations. An almost flat dispersion ($\Delta E=0.2$ meV) of the *single*-triplet mode at around 3.0 meV provides the direct evidence for the extremely localized nature in the orthogonal dimer system. In contrast to the *single*-triplet case, the $Q$ dependence of the *two*-triplets excitations displays more dispersive behavior originating from the correlation of the triplets. Moreover, a broad magnetic excitation at around 9 meV is observed for the first time, which is discussed in terms of correlated *three* or multi-triplets excitations.

Inelastic neutron scattering experiments were carried out on ISSP-PONTA spectrometer installed at 5G beam port of the Japan Research Reactor 3M (JRR-3M) in Japan Atomic Energy Research Institute, Tokai Establishment. The spectrometer was operated in the unpolarized neutron mode with pyrolytic graphite (PG) monochromator and analyzer. The inelastic scans were performed with fixed final energy $E_f=14.7$ meV ($k_f=2.67$ Å$^{-1}$) and horizontal collimations of open(40')-40'-sample-80'-80'. A PG filter was placed after the sample to suppress the higher order contaminations. In order to minimize the neutron absorption by the natural abundance of $^{10}$B, $^{11}$B-enriched (99.6%) bulk single crystals of $SrCu_2(^{11}BO_3)_2$ were prepared by the traveling solvent floating zone method using $Li^{11}BO_2$ flux [11]. The neutron scattering sample consisted of two single crystals with a total volume of 1.5 cm$^3$ aligned within 20' and it was oriented with its *a*- and *b*-axes in the scattering plane.

For convenience, we use a Brillouin zone at ($h$, $k$, 0) plane as given in Fig. 1, where we do not distinguish between the dimers lying along [1, 1] and [1, -1]. In this case, the space lattice vectors are transferred to $a_m=(a-b)/2$ and $b_m=(a+b)/2$, which correspond to the reciprocal space given by the lattice vectors $a_m^*=a^*-b^*$ and $b_m^*=a^*+b^*$. Shown in Fig. 2(a) are typical energy scans obtained at 1.7 K and 24 K for a scattering vector of $Q=(2, 0)$. An abrupt growth in intensity below 2 meV is due to the elastic incoherent scattering from the sample. The 1.7 K spectrum consists of three peaks centered at 3.0 meV, 5.0 meV and 9.7 meV (transitions I, II and III, respectively), while that for 24 K no longer has any appreciable peak, indicating that all excitations are of magnetic origin. As already confirmed by other measurements [5-10], transition

I at $\omega$=3.0 meV is the excitation of a single triplet from the singlet ground state. As shown in Fig. 2(b), the peak intensities of transitions I-III start to decrease with increasing temperature and disappear at about 13 K. We can derive from this fact that transitions II and III arise from multi-triplets excitations, which will be discussed in details later. It would be important to point out that the 3 meV mode loses all its intensity at 13 K, only one-third of spin gap energy (34 K), which is not characteristic of most gapped systems [12-16]. We suppose this anomalous temperature dependence indicates the strong spin frustration as well as the quantum effect of some kind in this orthogonal dimer spin system. A similar behavior has been also seen in Raman scattering study, in which several well-resolved modes for the multi-triplets excitations below 4.2 K are consiberably damped already at 7 K [8].

The $Q$ dependence of transition I was measured at 1.7 K along the lines A, B and C, and that of transitions II and III along D, E and F indicated by arrows in Fig. 1(a). It was found that, independent of $Q$, the obtained profile for transition I at 1.7 K is resolution limited, whereas the transitions II and III show intrinsic line widths. Accordingly, the 1.7 K profile was fitted to a combination of a delta function for transition I and two damped harmonic oscillators for transitions II and III, convoluted with the instrumental Gaussian resolution. In spite of large instrumental resolution about 1.3 meV for transition I, the excitation energies can be determined to an accuracy of 0.1 meV, less than 10 % of the instrumental resolution since the scattering intensity is strong enough. It is noted that the solid line in Fig. 2(a) includes the temperature independent term of incoherent scattering around energy zero.

The dispersion relation of band I is shown in Fig. 3. Most importantly, the excitation energies are almost $Q$ independent. Namely, the magnitude of the dispersion, the difference between the maximum and minimum of the excitation energy, is $\Delta E$=0.2 meV at most. This width is extremely small in contrast to conventional low-dimensional quantum spin systems. Experimentally, strong dispersions of the single-triplet excitations were observed in the 1D spin-Peierls material $CuGeO_3$ ($\Delta E$=14 meV) along the chain axis [12], and in the 2D plaquette system $CaV_4O_9$ ($\Delta E$=7 meV) parallel to the plaquette plane [13]. On the contrary, well-isolated clusters of exchange-coupled paramagnetic ions, where intercluster interactions are rarely important, evidently exhibit flat dispersions in the spin excitation spectrum. This has been experimentally shown by the neutron scattering measurements on the isolated dimer systems $Cs_3Cr_2Br_9$ ($\Delta E$=1.8 meV) [14] and $BaCuSi_2O_6$ ($\Delta E$=0.7 meV) [15], and the four-spin system $Cu_2PO_4$ ($\Delta E$=3 meV) [16]. In $SrCu_2(BO_3)_2$, the physical situation is completely different because the dimers within the ($a$, $b$) plane are not isolated but strongly interacting and moreover the system is located in the vicinity of the Néel ordered state.

As theoretically shown by Miyahara and Ueda [2], the key to the dispersionless band in $SrCu_2(BO_3)_2$ lies not in the spatial isolation of the spin clusters but in the orthogonality of the neighboring dimers: They proved that a hopping of the single triplet from one site to another within each plane is possible only from the sixth order in the perturbation calculations, leading to an exceedingly weak dispersion of band I. As mentioned above, the appearance of the quantized plateaux in the magnetization curve [2, 5] and the multiple magnetic resonances in ESR [7] indicate the localized character of *multi*-triplets excitations. Our neutron scattering study, however, provides for the first time a direct proof of the significant constraint on the *single*-triplet excitations. Recent NMR measurement by Kodama *et al.* [10] has also disclosed evidence to support the localization of the *single* triplet.

Let us take a closer look at the $Q$-dependence of band I, in Fig. 3. The dispersion curve reaches a maximum (3.10 meV) at the so-called ($\pi$, 0), a second maximum (3.00 meV) at ($\pi/2$, $\pi/2$), and a minimum (2.90 meV) at (0, 0) and the equivalent point ($\pi$, $\pi$). Using the series expansion method up to the fifteenth order, Weihong, Hamer and Oitmaa obtained the single-triplet excitation spectrum [4]. They argued that the bandwidth is quite small as anticipated, but increases as $J'/J$ approaches $(J'/J)_c$. For a comparison, the calculated dispersion curve is shown by the broken line in Fig. 3, where the exchange energies $J$=100 K and $J'$=68 K are assumed. From the qualitative point of view, the calculated $Q$ dependence of the excitation energies nicely reproduces our experimental data. However, the calculated bandwidth of 0.7 meV is (relatively narrow but) much wider than the observed one. In addition to this discrepancy, their theory yields a slightly smaller value of $\Delta$=2.0 meV at (0, 0) and ($\pi$, $\pi$). A possible explanation for the quantitative discrepancies is that the perturbative approach becomes no longer appropriate when we discuss the phenomena of the system near the critical boundary $(J'/J)_c$. Another possible explanation could be the three-dimensional effect, i.e., the existence of the interlayer coupling, which also could lead to a slightly different extimation of $J$ and $J'$.

We now discuss the transition II which occurs at energy transfer of about 5 meV. This excitation has been already observed by ESR [7] and Raman scattering [8] (at 4.7 meV), and the magnetic field dependence of the ESR frequency has identified this mode as the second triplet state [7]. Transition II is understood based on two triplets coupled by $J_t(>0)$: If $J_t$=0, the corresponding transition occurs only at $2\Delta$=6.0 meV. But when $J_t$ is finite, one expects a separation of the transition into three energy levels at $E_0=2\Delta-2J_t$ for the singlet state, $E_1=2\Delta-J_t$ for the triplet state, and $E_2=2\Delta+J_t$ for the quintet state. Using $E_1$=4.7 meV [7, 8] and $E_0$=3.7 meV [8], $J_t$ is determined to be 1.1-1.3 meV.

Figure 3 shows that the $Q$ dependence of band II is qualitatively identical to that of band I. The striking difference between bands I and II is that the latter shows more dispersive behavior with a bandwidth of 1.5 meV. This observation may indicate that the propagation of correlated *two* triplets is much easier than that of the *single* triplet. Very recent preliminary consideration of the two-triplets excitations by Miyahara and Ueda indicates that two triplets sitting on the nearest neighbor sites can propagate within the fourth-order perturbation calculations, leading to stronger dispersions of band II [17]. Considering the instrumental energy resolution, the full width at half maximum (FWHM) of band II for any $Q$ points is obtained to be approximately 0.7 meV. The finite width possibly indicates a spin continuum and/or a short lifetime of the coupled two triplets. One of the remaining problems regarding this mode is to determine the distance between a particular pair of the triplets. This may be possible if one analyses the $Q$ dependent intensity variation of this mode, which will be our future work.

The newly observed transition III which appears at 8-12 meV may be interpreted within the framework of correlated three triplets or more. Compared with the transition II, the observed profile of the transition III is considerably broader (see Figs. 2(a) and 3), which may result from a much wider spin continuum and/or a much shorter lifetime of the bound state. Or one might think the case of several excited states lying at nearly degenerate discrete energies. We suppose the excitations are also delocalized owing to the correlation of the multi-triplets as in the case of transition II though, at present, the very weak peak intensity and broad nature of transition III do not allow a reliable discussion of the dispersion relation.

Finally, we discuss the $Q$ dependence of the intensities of excitation I. The intensities at $T$=1.7 K and $E$=3.0 meV were measured over various $Q$-points. The data taken along three directions in the reciprocal lattice are shown in Fig. 4. The observed periodic intensity modulation was compared with the dynamical structure factor which can be obtained by calculating the transition probability from the singlet ground state to the lowest triplet excited state. For simplicity, let us consider a non-interacting pair of orthogonal dimers ($J'$=0) with the intradimer distance 2.905 Å. Then, the intensity at $Q$=($h$, $k$) is given by the incoherent superposition of the two single dimer contributions as $I(Q) \sin^2(Q\ d_1) + \sin^2(Q\ d_2) = \sin^2[0.717(h+k)]) + \sin^2[0.717(h-k)]$, where $2d_1$ and $2d_2$ are the vectors connecting individual spins of the two orthogonal dimers. The $Q$ dependence of the magnetic form factor is not included, because it does not affect the result considerably in the limited $Q$-range studied. Included in Fig. 4 are these theoretical curves, in qualitative agreement with the experimental

-6-

findings in spite of the unrealistic assumption about $J'$. This fact reflects the frustrating arrangement of the $J'$ bonds between the dimers.

In summary, we performed inelastic neutron scattering experiments on $SrCu_2(^{11}BO_3)_2$ to find three magnetic excitations. The lowest excitation from the ground state was observed at 3.0 meV, in agreement with previous measurements. We observed for the first time the almost dispersionless excitation in strongly correlated spin systems, originating not from the spatial isolation of the spin clusters but from the peculiar orthogonal dimer network . The second triplet mode at around 5 meV can be interpreted as the correlated two triplets excitations, but it is more dispersive than that of the 3 meV excitations. Likewise, the high-energy excitation in the energy range of 8-12 meV would arise from the correlated three or multi-triplets excitations.

The authors are grateful to H. Nojiri, P. Lemmens, M. Takigawa, S. Miyahara and K. Ueda for stimulating discussion. This work was supported by a Grant-in-Aid for Encouragement Young Scientists from The Ministry of Education, Science, Sports and Culture.

**References**

* Electronic address: kage@issp.u-tokyo.ac.jp

[1] See, e.g., M. Hase *et al*.: Phys. Rev. Lett. **70**, 3651 (1993); J. Darriet and J. P. Regnault, Solid State Commun. **86**, 409 (1993); M. Azuma et al., Phys. Rev. Lett. **73**, 2626 (1994); S. Taniguchi *et al*., J. Phys. Soc. Jpn. **64**, 2758 (1995).

[2] S. Miyahara and K. Ueda, Phys. Rev. Lett. **82**, 3701 (1999).

[3] B. S. Shastry and B. Sutherland, Physica **108B**, 1069 (1981).

[4] Z. Weihong, C. J. Hamer, and J. Oitmaa, Phys. Rev. B **60**, 6608 (1999).

[5] H. Kageyama, K. Yoshimura, R. Stern, N. V. Mushnikov, K. Onizuda, M. Kato, K. Kosuge, C. P. Slichter, T. Goto, and Y. Ueda, Phys. Rev. Lett. **82**, 3168 (1999).

[6] H. Kageyama, K. Onizuka, T. Yamauchi, Y. Ueda, S. Hane, H. Mitamura, T. Goto, K. Yoshimura, and K. Kosuge, J. Phys. Soc. Jpn. **68**, 1821 (1999).

[7] H. Nojiri, H. Kageyama, K. Onizuka, Y. Ueda, and M. Motokawa, J. Phys. Soc. Jpn. **68**, 2906 (1999).

[8] P. Lemmens, M. Grove, M. Fisher, G. Güntherodt, V. N. Kotov, H. Kageyama, K. Onizuka, and Y. Ueda, cond-mat/0003094.

[9] H. Kageyama, H. Suzuki, M. Nohara, K. Onizuka, H. Takagi, and Y. Ueda, to appear in Physica B.

[10] K. Kodama *et al*., unpublished.

[11] H. Kageyama, K. Onizuka, T. Yamauchi, and Y. Ueda, J. Crystal Growth **206**, 65 (1999).

[12] M. Nishi, O. Fujita, and J. Akimitsu, Phys. Rev. B **50**, 6508 (1994).

[13] K. Kodama, H. Harashina, H. Sasaki, Y. Kobayashi, M. Kasai, S. Taniguchi, Y. Yasui, M. Sato, K. Kakurai, T. Mori, and N. Nishi, J. Phys. Soc. Jpn. **66**, 793 (1997).

[14] B. Leuenberger, A. Stebler, H. U. Güdel, A. Furrer, R. Feile, and J. K. Kjems, Phys. Rev. B **30**, 6300 (1984).

[15] Y. Sasago, K. Uchinokura, A. Zheludev, and G. Shirane, Phys. Rev. B **55**, 8357 (1997).

[16] M. Hase, K. M. S. Etheredge, S-J. Hwu, K. Hirota, and G. Shirane, Phys. Rev. B **56**, 3231 (1997).

[17] S. Miyahara and K. Ueda, private communications.

Fig. 1: (a) The reciprocal lattice ($h$, $k$, 0) of the nuclear unit cell. The Brillouin zone used in this study is given by the broken lines. Along the lines A-C and D-F, the dispersion relations of band I, and bands II and III, respectively, have been studied. (b) The fundamental unit of the orthogonal dimers.

Fig. 2: (a) Energy scans at $Q$=(2, 0) obtained at $T$=1.7 K (circles) and 24 K (triangles). The peaks are labeled at the bottom of the figure. The solid curve is the fit to the data, as described in the text. (b) The temperature variation of the normalized intensities at $E$=3.15 meV and $Q$=(1.5, 0.5) (circles), $E$=5.15 meV and $Q$=(2, 0) (crosses), and $E$=10.65 meV and $Q$=(2, 0) (triangles). The broken line is a guide to the eye.

Fig. 3: $Q$-dependence of the excitation energies of bands I, II and III obtained at 1.7 K. The arrows represent the energy resolutions of the instrument (FWHM). The solid curves are guides to the eye. The bars represent the intrinsic line width (FWHM) of bands II and III. The theoretical dispersion curve for the single-triplet excitations by Weihong *et al*. [4] with $J$=100 K and $J$'=68 K is given by the broken line.

Fig. 4: Observed and calculated $Q$-dependence of the scattering intensity for excitation I.

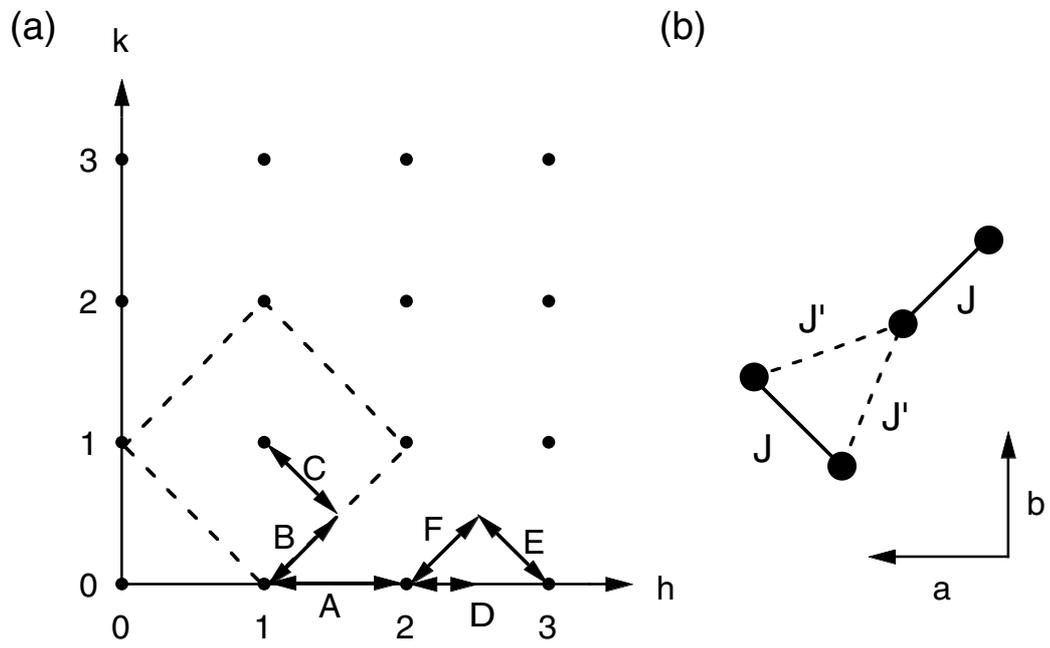

Fig. 1: Kageyama et al.

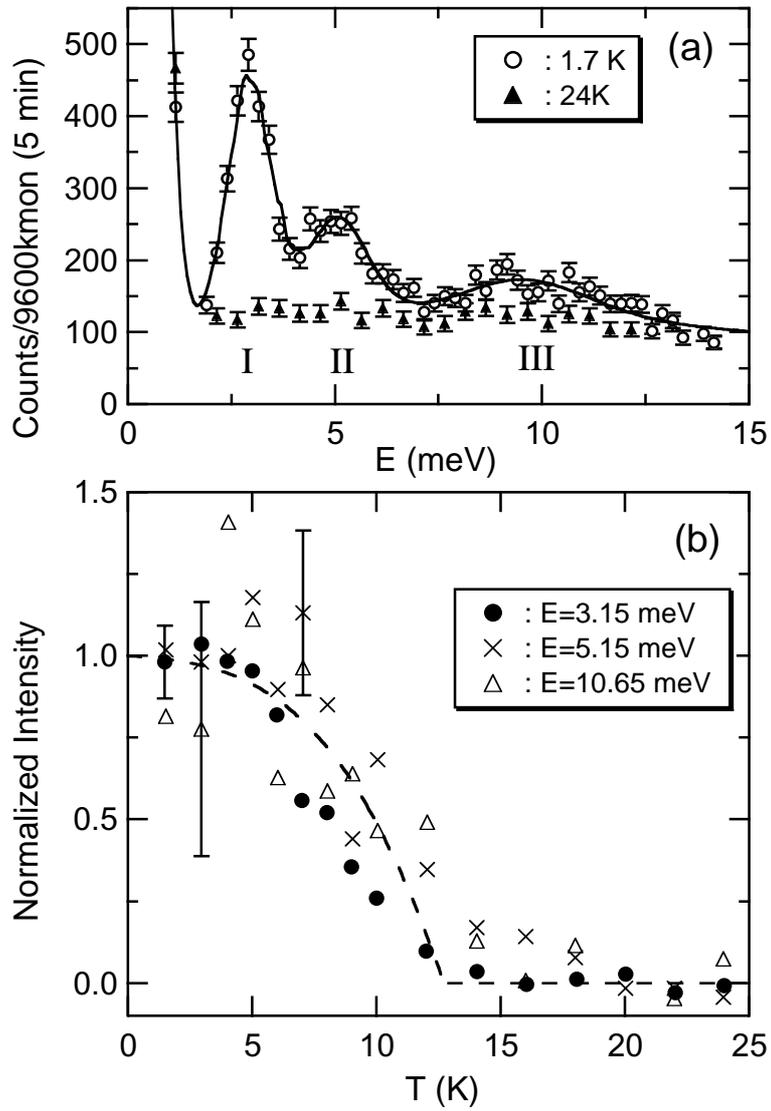

Fig. 2: Kageyama et al.

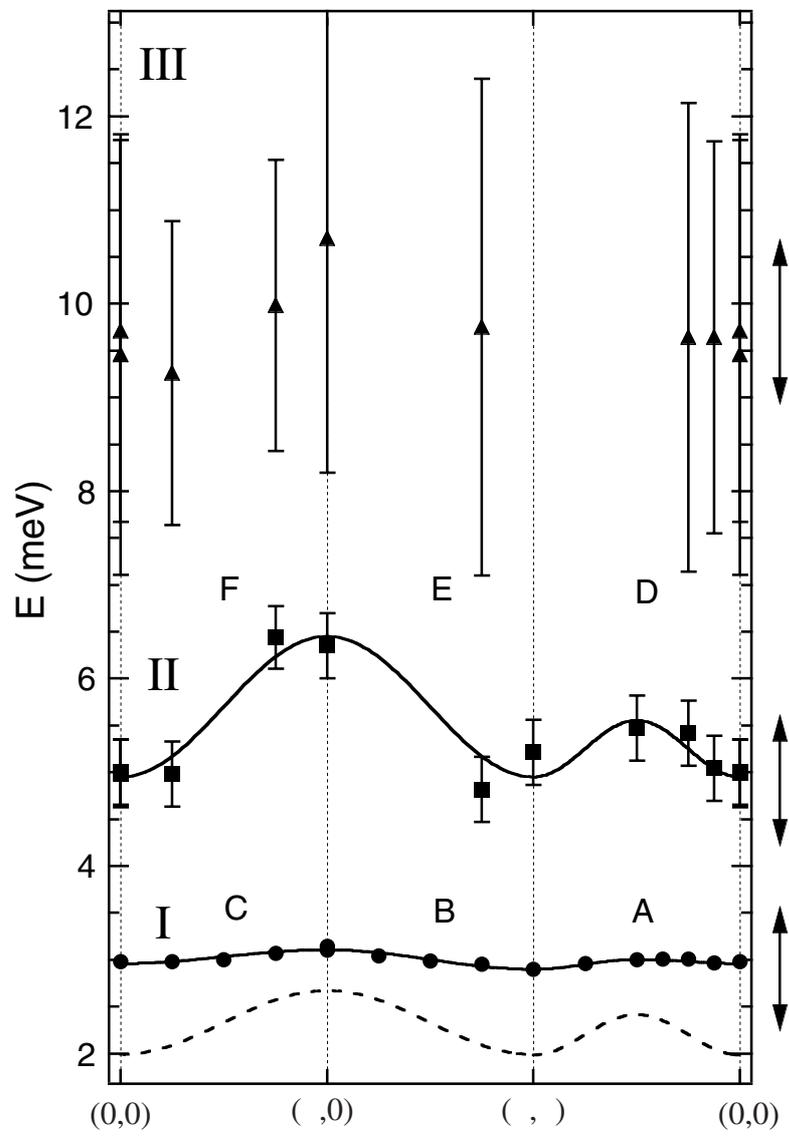

Fig. 3: Kageyama et al.

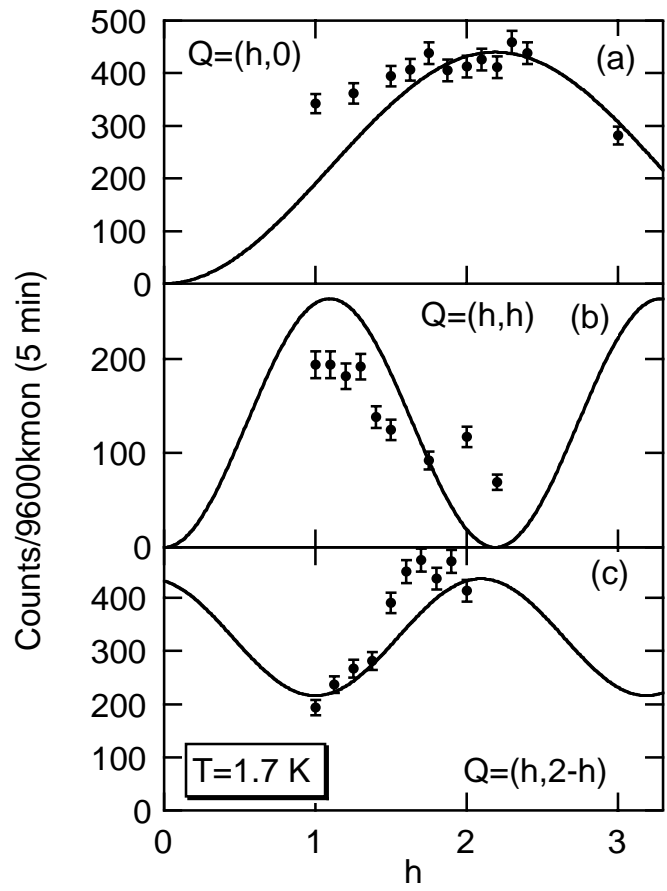

Fig. 4: Kageyama et al.